\documentclass[prl,aps,superscriptaddress,twocolumn,%showpacs,
 floatfix]{revtex4}
\usepackage{graphicx}
\usepackage{amssymb}
\usepackage[hypertex]{hyperref}
\usepackage{color}
\advance\belowdisplayskip by -0.2in
\advance\textfloatsep by -0.15in
\advance\abovecaptionskip by -0.15in
\advance\textheight by 0.1in
\begin{document}

\title{Quantum degenerate Bose-Fermi mixtures on 1-D optical lattices}

\author{Pinaki Sengupta}
\affiliation{Department of
    Physics \& Astronomy, University of Southern California, Los Angeles,
    California, 90089, USA} 
%\altaffiliation{Department of Physics, University of California,
%  Riverside, California 
%  92521, USA}
\author{Leonid P. Pryadko}
\affiliation{Department of Physics, University of California,
  Riverside, California 
  92521, USA}
                                                                              
\date{\today}

%%%%%%%%%%%%%%%%%%%%%%%%%%%%%%%%%%%%%%%%%%%%%%%%%%%%%%%%%%%%%%%%%%%
\begin{abstract}
We combine model mapping, exact spectral bounds, and a
  quantum Monte Carlo method to study the ground state phases of a
  mixture of ultracold bosons and spin-polarized fermions in a
  one-dimensional optical lattice.  The exact boundary of the
  boson-demixing transition is obtained from the Bethe Ansatz solution
  of the standard Hubbard model.  We prove that along a symmetry plane
  in the parameter space, the boson-fermion mixed phase is stable at
  all densities.  This phase is a two-component Luttinger liquid for
  weak couplings or for incommensurate total density, otherwise it has
  a charge gap but retains a gapless mode of mixture composition
  fluctuations. The static density correlations are studied in these
  two limits and shown to have markedly different features.
%%%%%%%%%%%%%%%%%%%%%%%%%%%%%%%%%%%%%%%%%%%%%%%%%%%%%%%%%%%%%%%%%%%
\end{abstract}


\maketitle

Recent advances in experiments with cold atoms in magneto-optical traps
have led to a series of exciting results. One area of emerging interest is 
the study of degenerate mixtures of bosons and fermions in various trap
and lattice geometries. While the original motivation for such studies 
was the sympathetic cooling of the fermions via interactions with bosons, 
such systems provide a unique opportunity to study experimentally 
quantum phases and associated transitions in a system of mixed quantum
statistics, which are rare in condensed matter systems. The absence of
defects and impurities as well as the ability to tune the interactions
between the particles will enable the exploration of various quantum
many-body phenomena. Stable boson-fermion mixtures have been realized 
with $^6$Li-$^7$Li\cite{li6li7}, $^{23}$Na-$^6$Li\cite{na23li6},
 $^{87}$Rb-$^{40}$K\cite{rb87k40} and $^6$Li-$^{87}$Rb\cite{li6rb87}.
On the theoretical front, the ground state properties
of boson-fermion mixtures -- in the continuum and on a lattice -- have 
been investigated using mean field theories\cite{molmer,albus,%
  das,lewenstein,miyakawa,akdeniz}, bosonization\cite{ho}, Bethe
ansatz (BA)\cite{imambekov} and numerical simulation of small
systems\cite{roth,takeyuchi}.

\begin{figure}[tbf]
\includegraphics[width=8.3cm]{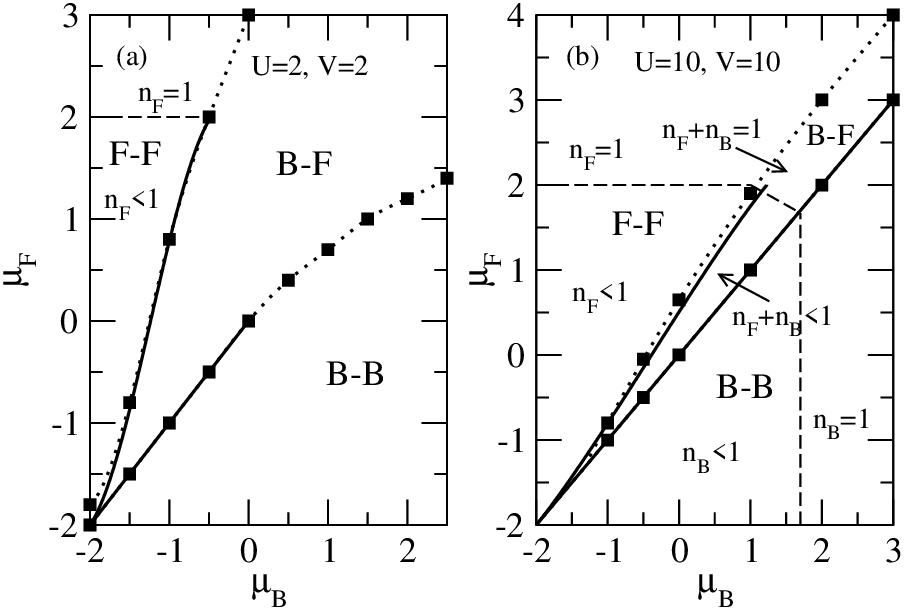}
\caption{Ground state phase diagram of the boson-fermion
  model~(\protect\ref{eq:model}).  Phases are purely bosonic (B-B),
  purely fermionic (F-F), and a uniform mixture of the two (B-F).  The
  filled squares joined by dotted lines to guide the eye represent
  the QMC results, while the solid lines denote the analytic phase
  boundaries.
  The dashed lines separate the commensurate (charge-gapped) and
  non-commensurate (gapless) sectors of the corresponding phases.  In
  the weak coupling limit, the B-F and B-B phases do not have a charge
  gap.}
\label{fig:phases}
\end{figure}

In this work we study the nature of the ground state of a
boson-fermion mixture on a 1D optical lattice, concentrating
specifically on the special case of equal hopping coefficients for
bosons and fermions, $t_B=t_F$ in Eq.~(\ref{eq:model}).  We construct
mappings to several known models in appropriate limits and use known
analytic results from these to supplement our numerical results.  In
addition to exploring the unique quantum demixing transitions, we have
investigated the nature of the ground state in the homogeneous (mixed)
phase at different interaction strengths and densities.  Our results
are summarized in Fig.~\ref{fig:phases}.  At large $\mu_F$ ($\mu_B$),
the ground state is purely fermionic (bosonic) with $n_B (n_F)=0$,
whereas a thermodynamically stable mixed phase (B-F) with $n_B,n_F>0$
exists over a finite range of parameters around $\mu_B\approx \mu_F$.
The exact analytic boundary between the pure fermion F-F and B-F
phases obtained from the solution of Eqs.~(\ref{eq:full-polarization})
and (\ref{eq:chem-potential}) is also shown.  The agreement between
the two approaches is excellent in the weak coupling limit, whereas
the B-F phase is overestimated in the numerical data at strong
coupling.  
The extent of the mixed phase gets smaller at low densities, especially
at large $U$ and $V$.  For the  case $U=V$, we show that even
at very strong coupling, there is no direct transition between pure
bosonic and fermionic phases.  A mixed phase is
always stable, contrary to the mean field results\cite{das}.  At weak
couplings, the mixed phase is dominant and one needs large differences
in the chemical potentials for the two components to observe demixing,
consistent with the bosonization results\cite{ho}.

A mixture of interacting bosons and spinless fermions in an optical
lattice is well described by the one-band Bose-Fermi Hubbard model
(BFHM),
\begin{eqnarray}
  H&=&\sum_{i=1}^N\Bigl[-t_B(b_{i+1}^{\dagger}b_i + h.c.)
  -t_F (f_{i+1}^{\dagger}f_i + h.c.) \nonumber\\ 
  & &\hskip-5mm + {U\over 2}n_i^B(n_i^B - 1) + V 
  n_i^Bn_i^F - \sum_{\alpha=B,F}\mu_\alpha n_i^\alpha\Bigr ].
\label{eq:model}
\end{eqnarray}
Here,  $U$ and $V$ parametrize the on-site boson-boson and boson-fermion
interactions; we assume $U,V>0$. 
The chemical potentials $\mu_B$ and $\mu_F$ couple to the
densities of the individual components, $n_B$ and $n_F$. In the
experiments, there is 
also a trapping potential that confines the particles to a certain
region of the lattice.
As a simplification, we have ignored such a confining potential. The
present description is applicable in the central part of a system with
a shallow trapping potential. In the discussion of the numerical results, 
we assume $t_B=t_F=t=1$ which sets the energy scale.  All 
other parameters are (implicitly) expressed in terms of $t$. 

An insight into the model~(\ref{eq:model}) can be gained by studying
the limiting cases.  For $V=0$, the boson and fermion sectors are
decoupled.  The fermions are non-interacting; they fill a band with
momenta $|k|<k_F=\pi n_F$.  The fermion sector has a 
commensurate density and an associated band gap for $|\mu_F|>2 t_F$;
otherwise it is gapless.  The boson sector is described by the
Bose-Hubbard model\cite{jaksch}.  At incommensurate densities $n_B$,
the ground state is a gapless superfluid (SF) for any $U>0$; 
at commensurate densities, there is an SF to
Mott insulator transition at $U\approx3.5t_B$\cite{monien}.  Finally,
the boson sector is trivially an insulator at zero density, $n_B=0$, at 
$\mu_B<-2t_B$.

In the hard-core limit $U\to\infty$, the bosons can be mapped to a
second ``flavor'' of fermions.  Then, for $t_B=t_F$, the resulting
model can be rewritten as the usual Hubbard model, where the fermion
flavor serves as the (pseudo)spin index.  The effective external
magnetic field is $\tilde h=\mu_F-\mu_B$, the Hubbard coupling
constant is $\tilde U\equiv V$, the hopping $\tilde t=t_B$, and the
chemical potential $\tilde \mu=(\mu_B+\mu_F)/2$ couples to the density $n=
n_B+n_F$.  This model is
exactly solvable\cite{Lieb-Wu-1968}.  Most importantly, there is no
charge or spin gap away from the commensurate density, $0<n<1$.  Given
the density $n$, the Hubbard fermions reach full spin 
polarization at the field\cite{Frahm-Korepin-1990}
\begin{equation}
  \label{eq:full-polarization}
  \tilde h_c(n,\tilde U)
  = {\tilde U\over  \pi}\int _{0}^{\pi n}\!\! dk\, \cos k \,
  {\cos k-
  \cos \pi n\over (\tilde U/4\tilde t ) ^2 +\sin^2 k}.
\end{equation}
At this point one of the Luttinger components disappears, we are left
with only one kind of the fermions.  The corresponding value of the
chemical potentials
\begin{equation}
\mu_F=-2 \tilde t \cos \pi n,\quad
\tilde\mu=\mu_F+\tilde h_c(n,\tilde
U)/2.\label{eq:chem-potential}
\end{equation}
The system acquires a gap at the commensurate density, $n=0$,$1$.  The
corresponding values of the critical magnetic field are $\tilde
h(0,\tilde U)=0$, $\tilde h(1,\tilde U)=(\tilde U^2+16
\tilde t^2)^{1/2}-\tilde U$. 

For the original BFHM~(\ref{eq:model}), the field $\tilde
h=\mp \tilde h_c(n,V)$ would correspond to the demixing transition, where
the density of either fermions or bosons turns to zero.  We also note
that in the latter case, $n_B=0$, the boson repulsion constant $U$ is
irrelevant.  Therefore, as long as the transition is continuous (as
expected for $V\lesssim2U$), the
boson demixing transition, $n_B=0$, is expected at
$\mu_B=\mu_F-\tilde h_c(n,V)$.  The corresponding curve on the
$\mu_F$--$\mu_B$ diagram [Fig.~\ref{fig:phases}] is given
parametrically in terms of $n=n_F$ by
Eqs.~(\ref{eq:full-polarization}), (\ref{eq:chem-potential}). 

To understand the fermion demixing transition at $n_F=0$, notice that in the
``symmetric'' case, $t_B=t_F$, $U=V$, and $\mu_B=\mu_H$, the
Hamiltonian~(\ref{eq:model}) commutes with the superoperator, $Q=\sum
f_i^\dagger b_i$, and its conjugate, $Q^\dagger$.  A ground state
wavefunction (WF) with $N$ bosons and no fermions, $\Psi_B^{(0)}$, can
be mapped to an eigenstate with $N-1$ bosons and $N_F=1$,
$\Psi_B^{(1)}\equiv N^{-1/2}\,Q\Psi_B^{(0)}$. The
normalization was found using the anticommutator,
$\{Q,Q^\dagger\}=\hat N_B+\hat N_F$.  While $\Psi_B^{(1)}$ may not be
the ground state, the degeneracy implies that adding fermions is
definitely favorable for $\mu_F>\mu_B$.  Combined with the local instability of the fermion
phase at the line Eqs.~(\ref{eq:full-polarization}),
(\ref{eq:chem-potential}), this gives a proof of thermodynamical
stability of the mixed phase just above the symmetry line. 

The same argument could also work the other way, as long as the ground
state with $N_F=1$ fermion, $\Psi_{B1}^{(0)}$, contains a fully
symmetrical component, so that the corresponding bosonic WF
$Q^\dagger\Psi_{B1}^{(0)}\neq0$.  This happens at strong coupling
(certainly at $n=1$ discussed below), as well as at small densities,
as can be seen by analyzing the energy of the continuum BA
solution\cite{imambekov}(a). In these two cases, the fermion demixing 
transition is indeed located precisely at the line $\mu_F=\mu_B$ 
(solid diagonal in Fig.~1).
 
Finally, for strong but non-infinite couplings, $U,V \gg t_B,t_F$, the
low-energy physics of the system is described by the effective
Hamiltonian which at $t_B=t_F$ is analogous to the anisotropic
$t$-$J$-$V$ or extended Hubbard model in external magnetic
field\cite{Bariev-1994,Kitazawa-2003}.  The model has an 
easy-axis anisotropy for $V>2U$, which implies a spin gap disappears
via a first-order ``spin-flop'' transition (the discontinuous demixing
transition is accompanied by phase separation).  At the commensurate
total filling, $n=1$, the system acquires a charge gap and is further
reduced to the exactly-solvable XXZ spin-chain with the parameters
$J^z=2(t_B^2+t_F^2)/V-t_B^2/U$, $J^\perp=-2t_B
t_F/V$\cite{Pollet-2005}, in an additional magnetic field so that 
$n_B=1$ at the symmetry line discussed above.  The operator $Q$ under
this mapping becomes a zero-$k$ spin-wave creation operator. 
Again, there is no spin gap for $|J^\perp|>J^z$ (the BF mixture is a
single-component Luttinger liquid), 
while for $|J^\perp|<J^z$ the system is fully gapped and the
corresponding fermion demixing transition is first-order. 

We have used the Stochastic Series expansion (SSE) quantum Monte Carlo
(QMC) method\cite{sse} to simulate the BFHM in the grand-canonical
ensemble on chains of length $L=8-128$.  Ground state results are
obtained by taking large enough values of the inverse temperature,
$\beta$, where $\beta=2L$ was sufficient. To characterize different
phases, we have studied the charge (superfluid) stiffness of the
fermions (bosons), $\rho_s^{\alpha}=\langle W_\alpha^2\rangle /2\beta
L$\cite{pollock}, $\alpha=F,B$, and the static susceptibility for the
density correlations,
\begin{equation}
\chi^{\alpha}({\bf q})={1\over L}\sum e^{i{\bf q}\cdot ({\bf r}_j-{\bf
    r}_i)}\int^{\beta}_0d\tau\langle n^{\alpha}_j(\tau)n^{\alpha}_i(0)
\rangle. 
\end{equation}

\begin{figure}
\includegraphics[width=8.3cm]{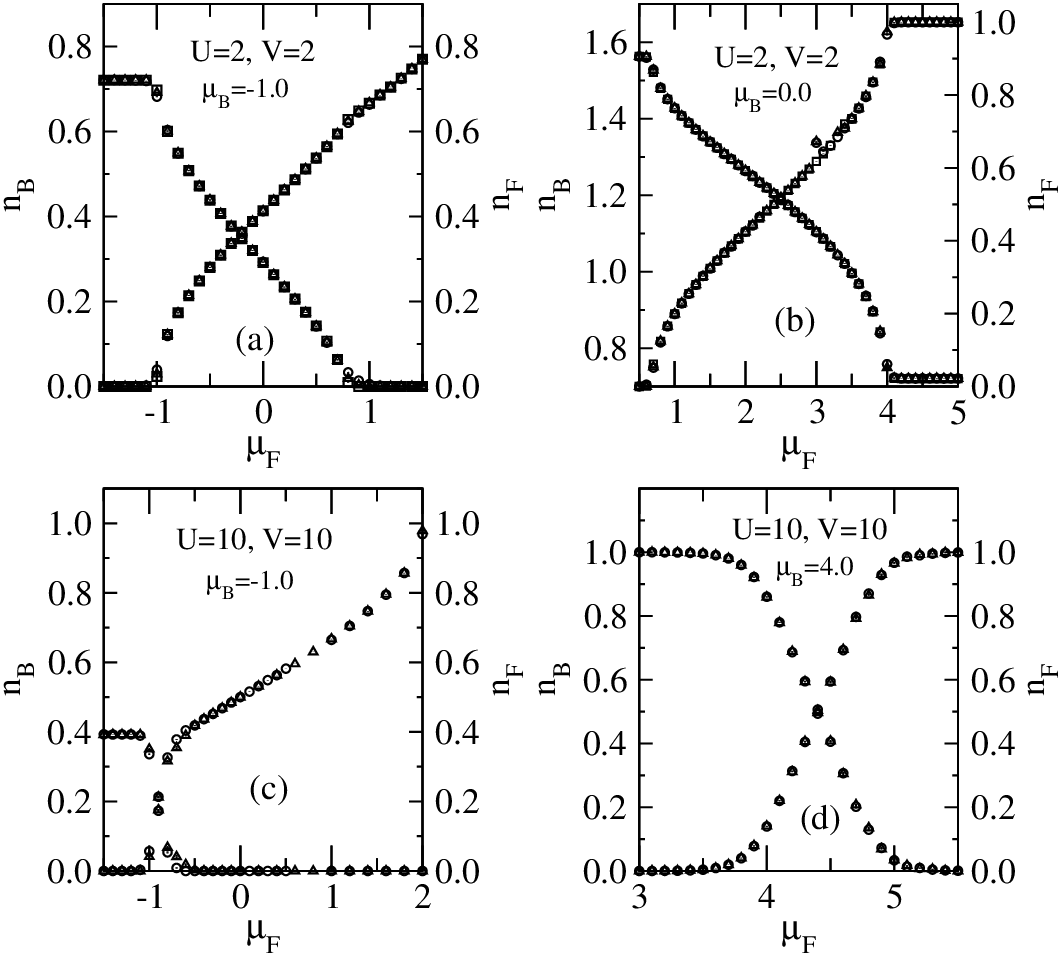}
\caption{The density of bosons and fermions as a function of the bare 
fermionic chemical potential, $\mu_F$.}
\label{fig:nbnf}
\end{figure}

We start the discussion of the numerical results by identifying the 
range of parameters which result in a thermodynamically stable 
homogeneous mixed phase.  We consider fixed (and equal) values of
$U$ and $V$ and vary $\mu_F$ and $\mu_B$. We have analyzed two cases
in detail -- one in the weak coupling regime ($U=V=2$), and the other
in the strong coupling regime ($U=V=10$).  Since SSE is formulated in
the grand canonical ensemble, one never observes a ground state with
spatially separated %pure bosonic and fermionic 
domains. Instead,
depending on the parameters of the Hamiltonian, the ground state will
be in one of three phases discussed above -- B-B, F-F, or B-F. Results
for $L=32-128$ are shown in Fig.~\ref{fig:nbnf}, where the bosonic and
fermionic densities are plotted as a function of varying $\mu_F$ at a
fixed value of $\mu_B$.  The upper(lower) panel shows the results for
the weak(strong) coupling regime. With increasing $\mu_F$ (at fixed
$\mu_B$), the ground state evolves continuously from B-B to B-F and
finally to F-F phase. The data (along with those for other values of
$\mu_B$) are combined to obtain the phase diagram Fig.~1.

Experimentally, the transitions can be observed in a large system with
fixed particle numbers if the trapping potentias seen by the two
species of particles vary differently.  For example, one can have a
shallow trapping potential for the bosons and a faster varying one for
the fermions. In that case, there will be purely fermionic domains at
the edges and a uniform mixture at the center of the trap.  Our data
show the transitions to be continuous, in contrast to the mean field
results, where the B-F to F-F transition is found to be
discontinuous\cite{das}. 

\begin{figure}
\includegraphics[width=8.3cm]{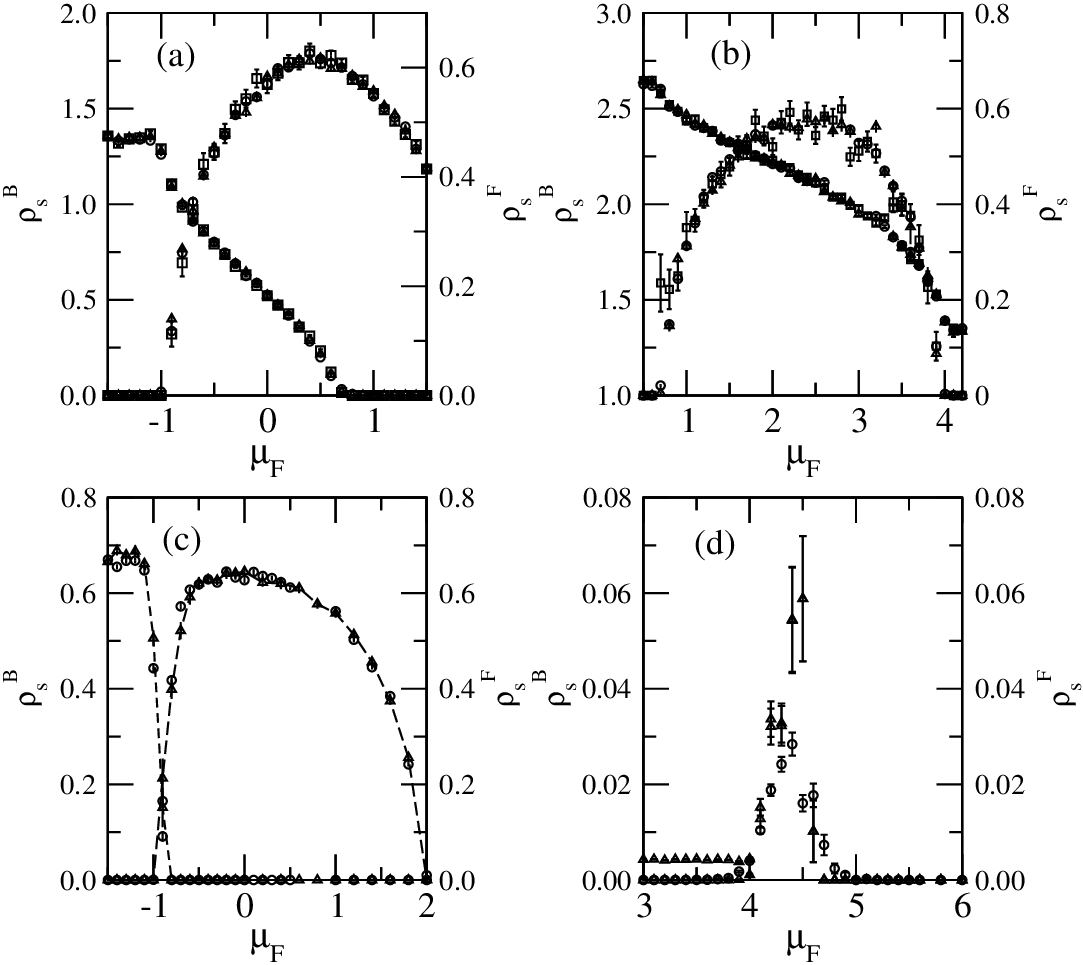}
\caption{Charge stiffness of the fermions ($\rho_s^F$) and the phase 
stiffness of the bosons ($\rho_s^B$) for the parameters in
Fig.~\protect\ref{fig:nbnf}.} 
\label{fig:rhos}
\end{figure}

Next we investigate the properties of the mixed phase in detail across
the parameter range explored in Fig.~\ref{fig:nbnf}.  We start with
the charge (phase) stiffness of the fermions (bosons) -- a finite
value implies a gapless spectrum for, and consequently, a Luttinger
liquid character of, the fermionic (bosonic) sector.  The results are
shown in Fig.~\ref{fig:rhos}. At weak couplings, $\rho_S^B$
($\rho_S^F$) is finite for any non-zero $n_B$ ($n_F$) (except
$\rho_S^F=0$ for $n_F=1$ as the fermions form a filled band).  In
particular both $\rho_S^B$ and $\rho_s^F$ are finite in the mixed
phase -- at low as well as high densities (or equivalently, bare
chemical potentials).  Hence the mixed phase in the weak coupling
limit is described by a two-component Luttinger liquid at all
densities.  

 In the strong coupling limit, the behavior is
markedly different at low and high densities.  At low densities
($n_F+n_B<1$) the stiffnesses are finite in the mixed phase, impying a
Luttinger liquid character of the ground state.  At large $\mu_{B/F}$, 
the system is always at the commensurate
filling, $n_F+n_B=1$.  In the pure phases, the stifnesses $\rho_S^B$
and $\rho_s^F$ vanish with the corresponding densities, $n_B$ and
$n_F$.  Also, $\rho_S^B$ vanishes in the B-B phase in the thermodynamic limit  
since the bosons are in the Mott insulating
phase at this $U$\cite{monien}.  In the mixed phase, the individual
stiffnesses have small, but finite, values\cite{note}. This suggests the
existence of a gapless mode, in contrast to previous  studies\cite{takeyuchi}.
The stiffness data and density correlations (shown below) support
the mapping to an equivalent XXZ chain, which at $U=V$ further reduces
to the gapless XY chain. The gapless mode consists of the $k=0$
spin-wave created by the superoperator $Q$ and corresponds to the
``pairing superfluidity'' in the BFHM explored in detail in 
Ref.~\onlinecite{Pollet-2005}.

\begin{figure}
\includegraphics[width=8.3cm]{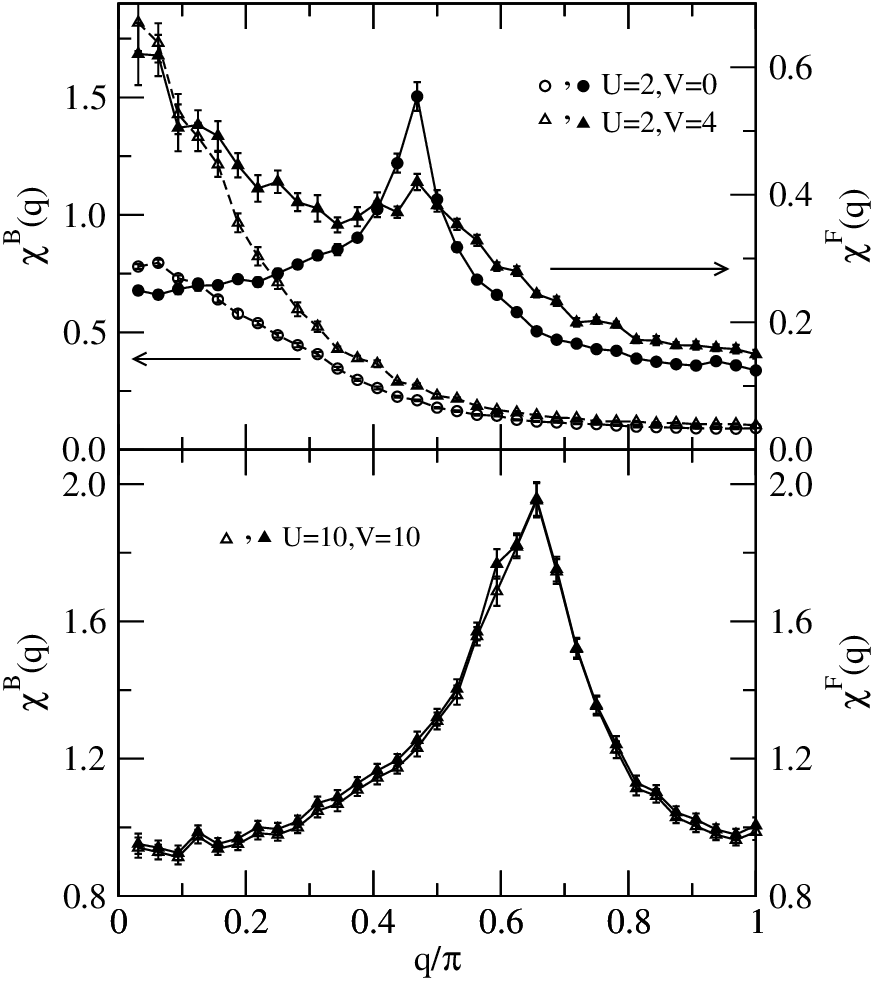}
\caption{(a) The static charge susceptibility for the fermions,
  $\chi_F(q)$,  
and bosons, $\chi_B(q)$, with varying inter-species
interaction strength, $V$. The peak at the Fermi wave vector is almost 
completely destroyed for $V=4$. The densities of the two species are 
$n_B\approx 0.17$ and $n_F\approx 0.24$. The open (filled) symbols and
dashed (solid) lines represent the bosonic (fermionic) data. 
(b) The suseptibilities in the
strong coupling limit at commensurate filling, $n_B\approx 0.67$ and 
$n_F\approx 0.33$.}
\label{fig:chi}
\end{figure}

To further characterize the nature of the mixed phase, we have studied
the static charge susceptibility [Fig.~\ref{fig:chi}] for the two components.
Once again we
find the results to be markedly different in the strong and weak
coupling regimes. Fig.~\ref{fig:chi}(a) shows the results of
increasing inter-species interaction on the static charge
susceptibility in the weak coupling regime. In the absence of
boson-fermion interaction, $\chi_F(q)$ shows the familiar peak at
twice the Fermi momentum (2$k_F$). As the inter-species repulsion is
increased, the peak at 2$k_F$ decreases and more weight is transfered
to lower momenta. Finally at large values of $V$, the peak at the
Fermi momentum is almost completely washed out implying that in this
regime the fermions no longer retain an independent character. On the
other hand, in the strong coupling limit, at commensurate filling
($n_B+n_F=1$), the results are diametrically opposite. The boson
density correlation develops a peak at twice the bosonic ``fermi
momentum'', 2$k_B=2\pi n_B$. In this limit of large $U$, the bosons
behave effectively as hard-core bosons whose diagonal correlations are
identical to those of spinless fermions.

To conclude, we have studied the ground state phases of the BFHM on a
1D optical lattice.  The analytical
expressions~(\ref{eq:full-polarization}), (\ref{eq:chem-potential})
for the boson-demixing boundary, constructed assuming a continuous
transition, are in an excellent agreement with the numerical results.
For the symmetrical interactions, $t_B=t_F$, $U_B=V_F$, the F-F phase
becomes locally unstable at $\mu_F=\mu_B+\tilde h_c(n,V)>\mu_B$
[Eq.~(\ref{eq:full-polarization})], any phase along the diagonal
$\mu_B=\mu_F$ is gapless to boson-fermion conversion, while the
fermion demixing transition happens strictly for $\mu_B\ge\mu_F$.
This establishes the stability of the mixed phase at all densities and
couplings.  
For strongly interacting bosons and fermions, the extent of the mixed
phase gets progressively narrower at low densities.  On the other hand,
for weakly interacting particles, the homogeneous phase is stabilized
over a wide range of chemical potentials.
Investigation of the properties of the B-F phase at $U=V$
reveals that at weak interactions, the mixed phase is a two-component
Luttinger liquid at all densities. For strong interactions, this
statement is valid at low densities, but breaks down at the
commensurate total density, $n_F+n_B=1$. Numerical data suggest a 
gapless mode for the ground state and strong coupling mapping
to the XXZ model predict that at $U=V$ this gapless mode consists 
of a $k=0$ spin-wave. In the BFHM, this corresponds to fluctuations 
of the mixture composition\cite{Pollet-2005}.  We have also explored 
the effects of 
boson-fermion interaction on the density correlations of the two
components and find that the results are dramatically different in the
weak and strong coupling limits.  These signatures should be readily
observable in experiments, and, if confirmed, would indicate the
emergence of novel many-body phases in these systems.

\acknowledgments We are grateful to L. Pollet, M. Troyer, and C. Varma
for useful discussions and for sharing their results before
publication (LP and MT). PS was supported by the DOE under grant
DE-FG02-05ER46240. Simulations were carried out in part at the HPC
Center at USC.

%%%%%%%%%%%%%%%%%%%%%%%%%%%%%%%%%%%%%%%%%%%%%%%%%%%%%%%%%%%%%%%%%%%

\end{document}